# Artificial Intelligence in Spanish Gastroenterology: high expectations, limited integration. A national survey

**Short title:** AI Adoption Gap in Gastroenterology

**Authors:** Javier Crespo[*] [1,4], Ana Enériz [2], Paula Iruzubieta [3], Fernando Carballo [4,5], Conrado Fernández Rodríguez [6,7], María Dolores Martín-Arranz[8,9], Federico Argüelles-Arias [10,11], Juan Turnes [2,4,12]

* Corresponding author:
   Email: javiercrespo1991@gmail.com

**Affiliations:**

([1]) Department of Medicine and Psychiatry, Faculty of Medicine. Clinical and Translational Research Group in Digestive Diseases, Valdecilla Research Institute (IDIVAL), Santander, Spain

([2]) IDARA Research Group, Galicia Sur Health Research Institute (IISGS), Vigo, Spain.

([3]) Digestive Diseases Service, Marqués de Valdecilla University Hospital. Clinical and Translational Research Group in Digestive Diseases, Valdecilla Research Institute (IDIVAL), Santander, Spain

([4]) MedicineAI, Madrid, Spain.

([5])University of Murcia, Murcia, Spain.

([6])Service of Gastroenterology, Hospital Universitario Fundación Alcorcón

([7])Department of Medical Specialties and Public Health, Rey Juan Carlos University, Madrid, Spain.

([8]) Department of Gastroenterology of La Paz University Hospital. School of Medicine. Universidad Autónoma de Madrid.

([9])Hospital La Paz Institute for Health Research, La Paz Hospital, Madrid, Spain.

([10]) Digestive Diseases Service, Virgen Macarena University Hospital, Seville, Spain

([11]) Faculty of Medicine, University of Seville, Seville, Spain

([12]) Department of Gastroenterology and Hepatology, Pontevedra University Hospital Complex, Pontevedra, Spain.

**Word Count**: 2.249

**Abstract**

**Background**. Artificial intelligence (AI) has emerged as one of the most disruptive innovations in medicine. However, its degree of adoption within gastroenterology remains limited and poorly characterized. We aimed to examine knowledge, practical applications, perceived barriers, and expectations regarding AI among gastroenterology specialists in Spain.

**Methods**. We conducted a cross-sectional observational study using a structured online survey distributed by the Spanish Society of Digestive Pathology (SEPD) between February and April 2025. The questionnaire collected sociodemographic data, patterns of AI use, perceptions, barriers, educational needs, and institutional expectations. Descriptive statistics, association analyses, and multivariable models were applied.

**Results**. Of 2,857 eligible SEPD members, 283 gastroenterologists responded (response rate: 9.9%; mean age 44.6±9.7 years; 59% women). Although 87.5% acknowledged AI as a transformative tool for the specialty, only 60.2% (95% CI: 54.3–66.1%) reported using it, mostly outside institutional frameworks. Notably, 80.2% of users had initiated AI use within the past year. Independent predictors of frequent use (≥1/week) included previous training (OR=2.44; 95% CI: 1.22–4.88), employment in university hospitals (OR=2.14; 95% CI: 1.33–3.45), and younger age (OR=1.36 per 5-year decrease; 95% CI: 1.12–1.64). Main barriers were lack of training (61%), absence of institutional strategies (46%), and ethical concerns (50%). While 93.8% agreed that AI training programmes are necessary, only 18.4% had received any formal training.

**Discussion**. A substantial gap exists between the favorable perception of AI and its actual integration into clinical practice within Spanish gastroenterology. The rapid recent adoption, predominantly outside institutional frameworks, underscores the urgent need for accredited training programmes and governance standards. Scientific societies may play a central role in leading this transition.

**Introduction**

Artificial intelligence (AI) has rapidly become one of the most transformative technologies in modern medicine. Its capacity to proces s large volumes of data and recognize complex patterns has been shown to improve diagnostic accuracy, optimize clinical decision-making, and accelerate biomedical research (1,2). Within gastroenterology, several applications have already demonstrated high levels of efficacy. Computer-aided detection systems have achieved sensitivities of 93% and specificities of 89% for the identification of colorectal polyps (3); deep learning algorithms have matched the performance of expert endoscopists in detecting Barrett's esophagus (4); and in inflammatory bowel disease, AI has proven useful in enabling an objective evaluation of endoscopic activity (5). In hepatology, predictive models are enhancing risk stratification in patients with metabolic dysfunction-associated liver disease (6).

Despite this promise, clinical implementation remains fragmented. It is estimated that fewer than 20% of gastroenterologists regularly use AI in their practice (7). In Italy, for example, while 72% of specialists acknowledged its utility, only 18% applied it in routine care, citing lack of training, regulatory uncertainty, and ethical concerns as the main barriers (8).

The Spanish context adds unique challenges. The European Union's Artificial Intelligence Act classifies AI-driven healthcare systems as "high risk," requiring strict monitoring (9). Moreover, the decentralized structure of the national health system and the pivotal role of scientific societies in postgraduate training create a distinct framework. The advent of generative tools such as ChatGPT has broadened access but has also introduced risks associated with unsupervised use (10,11). International evidence highlights that the adoption of health technologies depends not only on technical performance but also on organizational, cultural, and educational factors (12). The NASSS framework (Non-adoption, Abandonment, Scale-up, Spread, Sustainability) (13) underscores technological complexity, institutional preparedness, and sociocultural context as critical determinants of integration. In gastroenterology, these aspects are particularly relevant given the invasive nature of many procedures and the need to preserve physician–patient trust.

Against this backdrop, the present study, promoted by the Spanish Society of Digestive Pathology (SEPD), aimed to characterize knowledge, use, and perceptions of AI among gastroenterology specialists in Spain. Specifically, our objectives were to: (1) describe patterns of use and compare them with international references; (2) identify barriers and facilitators within the national regulatory context; (3) analyze educational needs across professional profiles; and (4) define the expected role of scientific societies in the digital transition.

**Materials and Methods**

We present a cross-sectional observational study using an anonymous electronic survey between February and April 2025, in accordance with the STROBE guidelines for observational studies (14). The target population comprised gastroenterology specialists (GS) practicing in Spain. The study population was defined as SEPD members with an active email address (N=2857). Sampling was non-probabilistic, based on voluntary participation. To ensure feasibility of the analyses, we aimed to collect at least 250 valid responses.

The questionnaire was developed following a critical review of the literature on AI perception and use in medicine and gastroenterology. It was refined through expert consensus within the SEPD Digital Innovation and Continuing Education panel. The final survey included items on sociodemographic variables, practice setting, professional experience, AI usage patterns and attitudes, perceived barriers, training needs, and expectations regarding the role of scientific societies. Open-ended questions were incorporated to capture participants' narratives on opportunities, risks, and proposals for integrating AI into clinical practice (the complete questionnaire is available in Supplementary Material).

The survey was distributed via the SEPD platform, with biweekly reminders throughout the data collection period. Participation was voluntary, unpaid, and anonymous. Informed consent was obtained at the beginning of the questionnaire.This study was conducted in full compliance with the ethical principles of the Declaration of Helsinki. All participants were informed about the study's objectives and voluntarily consented to participate. The survey collected data from professionals and did not involve patient data or clinical intervention, thus not requiring formal approval from an Institutional Review

Board under local regulations. All personal data were processed in accordance with the General Data Protection Regulation (GDPR) (EU) 2016/679 and Spain's Organic Law 3/2018 on the Protection of Personal Data and Guarantee of Digital Rights (LOPDGDD). The data were anonymized prior to analysis to ensure the confidentiality and privacy of all respondents

**Statistical analysis**

Quantitative data were described as means with standard deviations or medians with interquartile ranges, depending on distribution. Categorical variables weresummarized as absolute frequencies and percentages. Associations between categorical variables were evaluated using $\chi^2$ or Fisher's exact test; comparisons of means were performed with Student's t-test or ANOVA, as appropriate. Independent predictors of frequent AI use were assessed through multivariable logistic regression models. Qualitative responses were analyzed using thematic coding. Principal component analysis (PCA) was performed to identify underlying dimensions of AI perception. The resulting factor scores were used as input variables for cluster analysis to classify respondents into distinct professional profiles. All analyses were performed using R version 4.3.2 and Python version 3.11.

**Results**

A total of 283 gastroenterology specialists (GS) from all Spanish regions completed the survey, representing an estimated response rate of 9.9% of the target population. The mean age was 44.6±9.7 years (range 27–65); 59% were women, and more than half (56.2%) worked in university hospitals (Table1). Geographic distribution was representative, with no significant differences in participation across regions ($\chi^2$=14.3; p=0.42).

As shown in Table 2, 60.2% of respondents (95% CI: 54.3–66.1%) reported using some form of AI. AI users were significantly younger than non-users (41.2±8.9 vs 49.7±9.2 years; mean difference 8.5; 95% CI: 6.4–10.6; p<0.001) and more frequently employed in university hospitals (61.8% vs 47.8%; difference 14%; 95% CI: 2.8–25.2%; p=0.021) (Table 1). Temporal analysis revealed a marked acceleration: only 0.7% had initiated AI use more than three years earlier, whereas 80.2% reported first use within the past year. Regarding frequency, 38.2% reported weekly use and 28.7% daily use (Table 2). The most used applications were automated imaging analysis (82.2%), text-based assistants for clinical reporting (80.6%), and clinical decision support systems (70.5%).

Overall perception of AI was strongly positive, with a mean score of 3.92±0.61 on the validated 5-point Likert scale. Using principal component analysis (PCA) of AI perception items followed by cluster analysis on the resulting component scores, we identified four professional profiles: (31.8%) user with higher scores on items related to trust and expected expansion; (23.7%) non users with higher scores on adoption intention; (25.4%) characterized by intermediate scores and lower readiness to adopt; and (19.1%) characterized by lower scores on perceived utility and intention to integrate AI into clinical practice In the multivariable model, three factors independently predicted frequent AI use (defined as ≥1 use per week): younger age (OR=1.36 per 5-year decrease; 95% CI: 1.12–1.64), employment in university hospitals (OR=2.14; 95% CI: 1.33–3.45), and previous training in AI (OR=2.44; 95% CI: 1.22–4.88). Neither sex nor subspecialty remained significant after adjustment (Table 3).

The most frequently reported barrier to the use of AI was lack of training (61%), following by absence of an institutional strategy (46%). A majority of respondents (93.8%) expressed agreement, combining those who strongly agreed (76.7%) and those who agreed (17.1%) with the statement. Only a minimal proportion (2.3%) indicated disagreement or strong disagreement.

Among respondents, 94.9% were willing to dedicate at least 1 hour per week to AI-related training, and 49.5% were willing to commit 2 hours or more. Training preferences varied significantly across professional profiles ($\chi^2$=34.7; p<0.001). In the subgroup of respondents who expressed willingness to engage in AI training, the main area of interest was clinical diagnosis (56.2%), with smaller proportions prioritising predictive outcome analysis (22.7%) and teaching/education (18.8%); other applications accounted for 2.3%.

Regarding institutional roles, 88% assigned primary responsibility for regulation and training to scientific societies. The most frequently selected roles were specialised training (37.1%), facilitating access to tools (12.0%), developing clinical guidelines and standards (7.8%), promoting research (7.1%), and ethical and institutional representation (5.6%).

A total of 186 open-text responses were analyzed (response rate 65.7%). Five main categories were identified in the qualitative analysis: structured training (91 responses, 48.9%), clinical validation (62, 33.3%), ethical–legal concerns (56, 30.1%), scepticism (38, 20.4%), and proposals for responsible implementation (44, 23.7%).

**Discussion**

This nationwide survey demonstrates that, in Spain, gastroenterology specialists (GS) perceive artificial intelligence (AI) very positively; however, its effective use remains limited and largely occurs outside institutional frameworks. Importantly, the definition of "frequent use" adopted in our study (at least once per week) may not necessarily reflect sustained adoption of AI in clinical practice. This nuance should be considered when comparing our findings with those of other countries, where adoption criteria are not always explicitly defined and may vary across surveys (12,15). Despite this caveat, our results align with those reported n international contexts.

60.2% of respondents (95% CI: 54.3–66.1%) reported using some form of AI, a proportion higher than those reported in other developed countries. In Italy, for example, a recent national survey revealed that while 72% of gastroenterologists valued AI, only 18% reported regular use (8). In the United Kingdom, 61% of specialists declared routine use, supported by national guidelines and safety standards that facilitated integration (16). Data from Germany and the United States place adoption rates between 50% and 60%, confirming that intellectual acceptance of AI remains broader than its actual application (17,18).

The factors associated with higher AI use in our study as younger age, employment in university hospitals, and previous training, mirror findings from the literature on technological adoption in medicine. A generational digital divide has been well documented, shaping acceptance of healthcare innovations (19), while academic settings have been shown to facilitate early adoption (20). Notably, our multivariable analyses confirmed that previous AI training doubled the likelihood of frequent use, providing strong evidence for the necessity of structured educational programs. This result is consistent with international experiences demonstrating that formal education significantly increases readiness for adoption (21). The identification of four distinct professional profiles: highlights the heterogeneity of professional attitudes and provides a practical framework to design segmented educational strategies. These findings resonate with diffusion of innovation theory, which distinguishes early adopters from more cautious groups (21).

The barriers reported by our respondents as lack of training, absence of institutional strategies, cost, and ethical dilemas; are consistent with findings from international reviews (22). Interestingly, active users identified additional barriers related to technical complexity and their immediate work environment, underscoring that practical experience reveals challenges not perceived by non-users. These findings are consistent with the NASSS framework, which describes how barriers to healthcare technology adoption evolve from initial technical issues toward broader organizational and ethical considerations as implementation progresses (13), and with diffusion of innovation theory (21).

Endoscopy represents one of the most mature and clinically validated domains for artificial intelligence implementation in gastroenterology and therefore provides a useful reference for interpreting our findings. Although evidence supports its clinical benefit and guideline endorsement (23), emerging concerns such as potential deskilling underscore the need for structured governance and training (24). This reflects the gap identified in our study between positive perceptions of AI and its limited, largely non-institutionalized clinical adoption.

Parallels with other medical specialties reinforce the generalizability of our findings. In radiology, most physicians express favorable attitudes toward AI but emphasize lack of education and regulatory frameworks as primary barriers (25). In dermatology, surveys reveal low confidence in AI use and a strong demand for dedicated training (26). In ophthalmology, acceptance is broadly positive, although variable across applications, with a recurring demand for transparency in algorithmic outputs (27). This convergence across disciplines underscores that the solutions inevitably rest on three pillars: education, regulation, and continuous evaluation.

Looking forward, the integration of AI in gastroenterology should be built upon three fundamental axes: the development of certified, modular training programs tailored to different professional profiles; institutional integration of AI under validated and auditable standards; and active leadership of scientific societies in drafting guidelines and accrediting centers. The role of scientific societies should extend beyond normative functions to encompass strategic planning, promotion of multicenter research, creation of ethical debate forums, and representation before national and international regulatory bodies (28). For the SEPD specifically, its involvement extends beyond facilitating this survey: these findings should guide its future policies on training, innovation, and AI governance, strengthening its position as a catalyst of digital transformation in Spanish gastroenterology. Furthermore, coordinated European surveys led by entities such as the United EuropeanGastroenterology (UEG) would be highly desirable to map AI adoption in a comprehensive and harmonized manner across Europe.

This study has limitations inherent to cross-sectional surveys: causality cannot be established, self-selection and self-reporting biases may influence results, and conclusions should be interpreted within the Spanish gastroenterology context. Another limitation could be the response rate, 9.9% of the target population; although low, it is similar to other professional surveys (25,26). Although sampling was non-probabilistic and response rate moderate, the large sample size, geographical coverage, and diversity of subspecialties provide valuable heterogeneity and a broad, albeit not fully representative, overview of the national situation.

In conclusion, AI adoption among gastroenterologists in Spain is in a transitional phase: while perceptions are highly favorable, clinical use remains partial and weakly institutionalized. Structured training, integration under clear regulatory frameworks, and active leadership by scientific societies are essential to transform expectations into safe, equitable, and clinically impactful adoption.

**Acknowledgments**

This manuscript made use of AI-assisted tools, specifically ChatGPT (GPT-5.1) and Gemini 2.5, to support drafting and language revision. All AI-generated content was produced under the supervision of the authors, who critically reviewed, verified, and revised the outputs, and who take full responsibility for the final content of the manuscript.

**Tables**

**Table 1. Sociodemographic and professional characteristics of participants (n = 283). Includes distribution by age, sex, work setting, hospital type, and main subspecialty.**

| Characteristic | Total (n=283) | AI users (n=170) | Non-users (n=113) | p-value |
|---|---|---|---|---|
| Age, mean (SD) | 44.6 (9.7) | 41.2 (8.9) | 49.7 (9.2) | <0.001[a] |
| Female gender, n (%) | 167 (59.0) | 100 (58.8) | 67 (59.3) | 0.934[b] |
| University hospital, n (%) | 159 (56.2) | 105 (61.8) | 54 (47.8) | 0.021[b] |
| >10 years experience, n (%) | 156 (55.1) | 76 (44.7) | 80 (70.8) | <0.001[b] |
| **Main Subspecialty** | | | | |
| – Hepatology | 110 (38.9) | 62 (36.5) | 48 (42.5) | |
| – Endoscopy | 95 (33.6) | 63 (37.1) | 32 (28.3) | |
| – Other | 78 (27.5) | 45 (26.4) | 33 (29.2) | |
| **AI Exposure and Research** | | | | |
| Previous AI training, n (%) | 52 (18.4) | 44 (25.9) | 8 (7.1) | <0.001[b] |
| Research participation, n (%) | 92 (32.5) | 68 (40.0) | 24 (21.2) | 0.001[b] |

[a]Student's t test; [b]χ² test.

SD: standard deviation; AI: artificial intelligence.

**Table 2: Frequency and Temporality of AI tool Usage**

| Usage Frequency | Percentage (%) |
|---|---|
| Global AI use (any tool) | 60.2 % |
| Weekly use | 38.2 % |
| Daily use | 28.7 % |
| Monthly use | 6.6 % |
| Occasional use | 26.5 % |
| **Temporality (Length of Use)** | **Percentage (%)** |
| Started <1 month ago | 8.1 % |
| Between 1–6 months | 25.0 % |
| Between 6–12 months | 47.1 % |
| Between 1–3 years | 19.1 % |
| More than 3 years | 0.7 % |

**Table 3. Predictors of frequent AI use (≥1 time per week): multivariable analysis. Adjusted odds ratios (OR), 95% confidence intervals (CI), and p-values are reported.**

| Variable | β | SE | Adjusted OR | 95% CI | p-value |
|---|---|---|---|---|---|
| **Age (per 5 years younger)** | 0.307 | 0.098 | 1.36 | 1.12–1.64 | **0.002** |
| **University hospital** | 0.761 | 0.243 | 2.14 | 1.33–3.45 | **0.002** |
| **Previous AI training** | 0.892 | 0.354 | 2.44 | 1.22–4.88 | **0.012** |
| **Research participation** | 0.637 | 0.289 | 1.89 | 1.07–3.34 | **0.028** |
| **High perceived benefit** | 0.837 | 0.247 | 2.31 | 1.42–3.76 | **0.001** |
| High ethical concerns | −0.174 | 0.339 | 0.84 | 0.43–1.63 | 0.607 |
| Female gender | −0.083 | 0.235 | 0.92 | 0.58–1.46 | 0.723 |
| Constant | −2.147 | 0.486 | – | – | <0.001 |

Model performance: $\chi^2$=89.34, p<0.001;

Nagelkerke $R^2$=0.38; ROC AUC=0.78 (95% CI: 0.72–0.84).

Hosmer–Lemeshow: $\chi^2$=6.84, p=0.55.

Key: SE: standard error; OR: odds ratio; CI: confidence interval.c Based on factor scores >75th percentile.

# Supplementary material

# Survey

## Part 1. Demographic and Professional Data

1. Age:
 ☐ <30 ☐ 30–40 ☐ 40–50 ☐ 50–60 ☐ >60

2. Sex:
 ☐ Male ☐ Female ☐ Prefer not to answer

3. Years of experience in gastroenterology:
 ☐ Resident ☐ <5 years ☐ 5–10 years ☐ 10–20 years ☐ >20 years

4. Type of workplace:
 ☐ Public hospital ☐ Private hospital ☐ Mixed (public + private) ☐ Other

5. Is your hospital a university hospital?
 ☐ Yes ☐ No

6. Region in Spain:
 Autonomous community (selection)

7. Main area of professional dedication:
 ☐ Hepatology ☐ Diagnostic/therapeutic endoscopy ☐ Inflammatory bowel disease ☐ Motility/functional disorders ☐ Digestive oncology ☐ Nutrition ☐ General gastroenterology ☐ In training ☐ Other

8. Have you previously received any formal training in artificial intelligence?
 ☐ Yes ☐ No

## Part 2. Perception of Artificial Intelligence in Gastroenterology

Please rate your level of agreement (1–5): 1 = Strongly disagree | 5 = Strongly agree

9. AI is relevant for improving clinical care in gastroenterology. ☐1 ☐2 ☐3 ☐4 ☐5

10. AI has the potential to transform gastroenterology. ☐1 ☐2 ☐3 ☐4 ☐5

11. AI has had, so far, a major impact on gastroenterology. ☐1 ☐2 ☐3 ☐4 ☐5

12. AI improves diagnostic accuracy in digestive diseases. ☐1 ☐2 ☐3 ☐4 ☐5

13. Artificial intelligence helps optimise clinical decision-making in gastroenterology and hepatology.
☐1 ☐2 ☐3 ☐4 ☐5

14. AI reduces the workload of healthcare professionals.
☐1 ☐2 ☐3 ☐4 ☐5

15. AI should be integrated into gastroenterology training and education.
☐1 ☐2 ☐3 ☐4 ☐5

16. AI can and should be a driver of research and/or innovation in gastroenterology.
☐1 ☐2 ☐3 ☐4 ☐5

17. Artificial intelligence will improve the quality of research in gastroenterology.
☐1 ☐2 ☐3 ☐4 ☐5

18. AI raises ethical and data privacy concerns.
☐1 ☐2 ☐3 ☐4 ☐5

19. AI may replace some of the tasks I currently perform.
☐1 ☐2 ☐3 ☐4 ☐5

20. AI will be essential in the development of predictive models for the progression of digestive diseases.
☐1 ☐2 ☐3 ☐4 ☐5

21. Automated analysis of histological images using artificial intelligence will become widespread in the near future.
☐1 ☐2 ☐3 ☐4 ☐5

22. The use of chatbots and virtual assistants will become widespread in clinical practice
☐1 ☐2 ☐3 ☐4 ☐5

23. Predictive medicine will rely on artificial intelligence–based analysis of omics data.
☐1 ☐2 ☐3 ☐4 ☐5

**Part 3. Use of Artificial Intelligence in Professional Practice**

24. Do you currently use AI in your professional practice?
   ☐ Frequently  ☐ Occasionally  ☐ No, but I would like to  ☐ No, and I do not consider it necessary

25. In which areas do you think artificial intelligence could help you in your work in the short term?(Select all that apply)
☐ Drafting documents or clinical protocols
☐ Writing or supporting clinical records,
☐ Patient management,
☐ Image interpretation

26. If you do not use artificial intelligence, what is the main reason?
(Select all that apply)
☐ I do not have access to these tools
☐ I do not consider them useful
☐ Lack of training in artificial intelligence
☐ Lack of financial resources to access the technology
☐ There is no clear need in my current practice
☐ Lack of institutional or workplace support
☐ Concerns about data privacy

27. How long have you been using AI tools?
   ☐ <1 month  ☐ <6 months  ☐ 6–12 months  ☐ 1–3 years  ☐ >3 years

28. How often do you use AI tools?
   ☐ Daily  ☐ Weekly  ☐ Monthly  ☐ Occasionally  ☐ Never

29. For which activities do you use AI? (multiple choice)
☐ Diagnostic support
☐ Image interpretation
☐ Risk prediction
☐ Clinical decision support
☐ Education
☐ Innovation and new projects
☐ Clinical research

30. In which domains do you mainly apply AI?
   ☐ Clinical care  ☐ Education  ☐ Research  ☐ Innovation

Regarding the following questions, please rate your degree of use of artificial intelligence on a scale from 1 to 5, where:

1 = I do not have a formed opinion
2 = No, and I do not consider its use necessary
3 = No, but I would like to
4 = Yes, occasionally

5 = Yes, frequently

31. Have you used artificial intelligence for clinical care activities?
☐1 ☐2 ☐3 ☐4 ☐5

32. Have you used artificial intelligence for educational or teaching activities?
☐1 ☐2 ☐3 ☐4 ☐5

33. Have you used artificial intelligence for research activities?
☐1 ☐2 ☐3 ☐4 ☐5

34. How frequently do you use chatbots or virtual assistants?
☐1 ☐2 ☐3 ☐4 ☐5

35. How frequently do you use clinical decision support systems?
☐1 ☐2 ☐3 ☐4 ☐5

36. How frequently do you use automated image interpretation systems (e.g., endoscopy)?
☐1 ☐2 ☐3 ☐4 ☐5

37. Have you participated in AI-related research or innovation projects?
☐ Yes, actively ☐ Yes, as collaborator ☐ No, but interested ☐ No

## Part 4. Barriers and Facilitators for AI Adoption

38. What factors limit the adoption of AI in gastroenterology? (multiple choice)
☐ Lack of training ☐ Lack of institutional strategy ☐ Cost ☐ Ethical/privacy concerns ☐ Lack of evidence ☐ Technical complexity

39. Which measures would facilitate AI adoption? (multiple choice)
☐ Training programs ☐ Institutional support ☐ Clear regulations ☐ Integration into electronic health records ☐ Scientific society support

## Part 5. Training Needs in Artificial Intelligence

Please rate your level of agreement (1–5): ): 1 = Strongly disagree | 5 = Strongly agree

40. Specific AI training programs are necessary for gastroenterologists. ☐1 ☐2 ☐3 ☐4 ☐5

41. AI training should be considered a current priority. ☐1 ☐2 ☐3 ☐4 ☐5

42. I would be willing to participate in AI training programs. ☐1 ☐2 ☐3 ☐4 ☐5

43. How much time would you be willing to dedicate weekly to AI training?
☐ <1 hour ☐ 1–2 hours ☐ 2–4 hours ☐ >4 hours

44. What type of AI training would be most useful for you? (multiple choice):
☐ Practical workshops ☐ Online courses ☐ Webinars ☐ Applied clinical guides

45. What type of AI tools would you like to incorporate into your professional practice? (multiple choice)

☐ Clinical diagnosis   ☐ Predictive outcome analysis   ☐ Teaching and education   ☐ Other

**Part 6. Role of Scientific Societies in Artificial Intelligence**

Please rate your level of agreement (1–5): 1 = Strongly disagree | 5 = Strongly agree

46. Scientific societies should play a role in AI implementation.  ☐1 ☐2 ☐3 ☐4 ☐5

47. Scientific societies in gastroenterology are currently playing an adequate role in relation to AI implementation.  ☐1 ☐2 ☐3 ☐4 ☐5

48. Scientific societies should include AI training programmes within their regular activities.  ☐1 ☐2 ☐3 ☐4 ☐5

49. Scientific societies should prioritise AI in their strategic plans.  ☐1 ☐2 ☐3 ☐4 ☐5

50. Scientific societies should lead AI regulation in gastroenterology.  ☐1 ☐2 ☐3 ☐4 ☐5

51. What do you consider should be the main objective of scientific societies regarding AI? (multiple choice)
   ☐ Promote training  ☐ Facilitate access to AI tools  ☐ Develop and publish clinical guidelines  ☐ Promoting research  ☐ Represent regulatory and ethical interests